\documentclass[aps,preprint,amsmath,amssymb,nofootinbib]{revtex4}
\usepackage{graphicx}
\usepackage{epstopdf}

\begin{document} %
%%%%%%%%%%%%%%%%%%

%%%%%%%%%%%%%%%%%%%%%%%%%%%%%%%%%%%%%%%%%%%%%%%%%%%%%%%%%%%%%%%%%%%%%

\title{Probe of the Randall-Sundrum-like model with the small curvature via light-by-light scattering at the LHC}

\author{S.C. \.{I}nan}
\email[]{sceminan@cumhuriyet.edu.tr} \affiliation{Department of
Physics, Sivas Cumhuriyet University, 58140, Sivas, Turkey}
%%%%%%%%%%%%%%%%%
\author{A.V. Kisselev}
\email[]{alexandre.kisselev@ihep.ru} \affiliation{A.A. Logunov
Institute for High Energy Physics, NRC ``Kurchatov Institute'',
142281, Protvino, Russian Federation}

\begin{abstract}
The LHC possibilities to constrain the parameters of the
Randall-Sundrum--like model with one warped extra dimension and
small curvature through the diphoton production in the
photon-induced process $pp \to p \gamma\gamma p \to p' \gamma\gamma
p'$ are investigated. Two acceptances of the forward detectors,
$0.015 < \xi< 0.15$ and $0.015 < \xi <0.5$, where $\xi$ is the
fractional proton momentum loss of the incident protons, are
considered. The sensitivity bounds on the 5-dimensional gravity
scale are obtained as a function of the LHC integrated luminosity.
\end{abstract}

\maketitle

%%%%%%%%%%%%%%%%%%%%%%%%%%%%%%%%%%%%%%%%%%%%%%%%%%%%%%%%%%%%%%%%%%%%%

%%%%%%%%%%%%%%%%%%%%%%%%
\section{Introduction} %
%%%%%%%%%%%%%%%%%%%%%%%%
\label{sec:intr}

The Standard Model (SM), which defines the fundamental particles and
their interactions at the electroweak energy scale, has been proven
in all experiments, including the LHC, which has been done so far.
Nevertheless, scientists are still in searching for solutions of
many problems that SM cannot give satisfactory a solution. The
hierarchy problem, which involves the large energy gap between the
electroweak scale and the gravity scale is one of these problems.
The most important answers to this unexplained phenomenon can be
given by beyond the SM theories which include additional dimensions.
Therefore, such models have attracted great attention in recent
years and many articles have been published in the literature.

\begin{figure}[htb]
\includegraphics[scale=0.8]{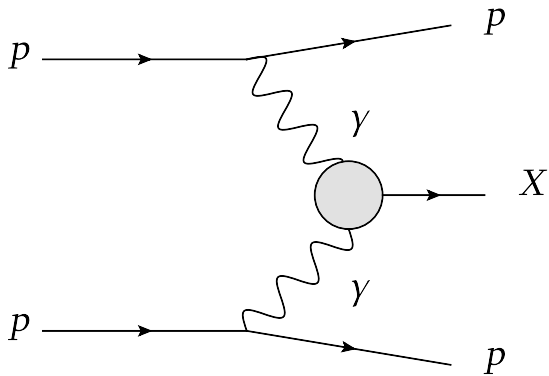}
\caption{Schematic diagram for the reaction $pp\to p \gamma\gamma p
\to pXp$. In our case, $X =  \gamma\gamma$.}
\label{fig:sch}
\end{figure}

At hadron colliders, inelastic collisions are generally performed
and their results are examined. However, the hadron colliders can
also be used as photon-photon, photon-proton colliders as applied in
the Tevatron \cite{cdf1,cdf2} and LHC
\cite{ch1,ch2,ath1,ath2,ath3,ath4}. The current results which are
found in these experiments are in agreement with theoretical
expectations. Specifically, the LHC experiments have shown that such
photon-induced processes are important for a search of new physics.
The most important advantage of the photon-induced process is that
it has a clean background. It's because that this process does not
include a lot of QCD originating backgrounds and uncertainties
resulting from proton dissociation into jets. All these backgrounds
make it difficult to identify the new physics signal beyond the SM.
The photon-photon collisions through the process  $pp \to
p\gamma\gamma p \to p'Xp'$ has very little background. Schematic
diagram for this collision is shown in Fig.(\ref{fig:sch}). As one
can see, both protons remain intact in this exclusive process.
%%%%%%%%%%%%
% Figure 1 %
%%%%%%%%%%%%

Examining photon-photon interactions at the LHC is possible thanks
to the plan prepared by the ATLAS Forward Physics (AFP)
Collaboration and joint CMS-TOTEM Precision Proton Spectrometer
(CT-PPS) \cite{afp1,afp2,afp3,totem}. These plans include forward
detectors which are placed symmetrically at a distance from the main
detectors. The forward detectors have charged particle trackers.
They can catch the intact protons after elastic photon emission in
the interval $\xi_{\min}<\xi<\xi_{\max}$ where $\xi$ is the
fractional proton momentum loss of the protons,
$\xi=(|\vec{p}|-|\vec{p}^{\,\,\prime}|)/|\vec{p}|$. Here $\vec{p}$ is the incoming proton
momentum and $\vec{p}^{\,\,\prime}$ is the momentum of the intact
scattered proton. The application of forward detectors to detect the
scattered protons is used to identify the collision kinematics and
consequently, photon-induced processes can be studied at the LHC.
Forward detectors should be installed closer to the main detectors
to achieve greater values of $\xi$.

AFP has the $0.0015 < \xi <0.15$, $0.015 < \xi <0.15$ detector
acceptance ranges. Similarly, detector acceptance ranges of the
CT-PPS are  $0.0015 < \xi< 0.5$, $0.1< \xi <0.5$. AFP includes two
types of studies. The first one is exploratory physics (anomalous
couplings between $\gamma$ and $Z$ or $W$ bosons, exclusive
production, etc.). The second one is the standard QCD physics
(double Pomeron exchange, exclusive production in the jet channel,
single diffraction, $\gamma\gamma$ physics, etc.). The main goals of
the CT-PPS experiment are the examination of the elastic
proton-proton interactions, the proton-proton total cross-section
and other diffractive processes. These charged particle detectors
unable to determine the almost all inelastic interactions in the
forward area. In this way, a very wide solid angle can be examined
with the support of the CMS detector. Also, the forward detectors
can be applied for precise studies \cite{ttm1, ttm2, ttm3}. Pile-up
events can occur as a result of such high luminosity and high energy
interactions. However, by using kinematics, timing constraints and
exclusivity conditions, these backgrounds can be extremely
restricted \cite{albrow,albrow1}. There are many phenomenological
papers in the literature which are based on the photon-induced
reactions at the LHC aimed at searching for physics beyond the SM
\cite{lhc2,lhc4,inanc,inan,bil,bil2,kok,inan2,gru,inanc2,epl,inanc3,bil4,inanc4,hao1,hao2,ins,kok2,fica,ficb,fic1,fic2,fic3,kok3}.

In the present paper we investigate the Randall-Sundrum-like model
scenario with the small curvature (the details are given in
Section~III) through the main process $pp \to p \gamma \gamma p \to
p' \gamma \gamma p'$ with the subprocess $\gamma \gamma \to \gamma
\gamma$ in this study for the $0.015<\xi<0.15$ and $0.015<\xi<0.5$.
First evidence for $\gamma \gamma \to \gamma \gamma$ scattering  was
observed by the ATLAS collaboration in high-energy ultra-peripheral
heavy ions collisions \cite{lbl1}. After that, the CMS collaboration
was reported for the same process \cite{lbl2}. Therefore, studies on
this process have gained more importance in recent times. Recently,
we have studied the photon-induced dimuon production at the LHC
\cite{Inan-Kisselev}. It is clear that any BSM scenario must be
checked in a variety of processes in order to find the most
appropriate one. As we will see below, the bounds on the main
parameter of the model for the diphoton production are better than
the bounds obtained in \cite{Inan-Kisselev}. Note that the process
going through the subprocess $\gamma\gamma \rightarrow
\gamma\gamma$, is known to be one of the most clean channels.

The processes contributing to the SM exclusive photon-photon
production consist of diagrams with charged fermions (leptons,
quarks), $W$ boson loop contributions and gluon loop diagrams. Also,
the interference terms of these processes should be taken into
account in order to obtain the whole SM cross section.

These processes have been examined in refs.
\cite{sm1,sm2,sm3,sm4,sm5}. QCD gluon loop contributions are
dominant at low energy regions whereas $W$ loop contributions
dominate at higher high energy regions. As shown in \cite{fic0} QCD
loop contribution can be neglected for the diphoton mass larger than
$200$ GeV. In our study, we have implemented the cut on the diphoton
mass of $200$ GeV, and therefore, we have omitted the QCD loop
contributions.

There are $16$ helicity amplitudes of the process ${\gamma\gamma \to
\gamma\gamma}$. However, if T-invariance, P-invariance and Bose
statistics are taken into consideration, the following relations are
obtained
\begin{eqnarray}
\label{sym} &&M_{++++}=M_{----}\; ; \quad M_{++--}=M_{--++} \; ;
\nonumber \\
&&M_{+-+-}=M_{-+-+}\; ; \quad M_{+--+}=M_{-++-} \; ;
\nonumber \\
&& M_{+++-}=M_{++-+}=M_{+-++}=M_{-+++}
\nonumber \\
&&=M_{---+}=M_{--+-}=M_{-+--}=M_{+---} \;.
\end{eqnarray}
\noindent With using there relations, the total matrix element takes
the form
\begin{eqnarray}
|M|^2=2|M_{++++}|^2+2|M_{++--}|^2+2|M_{+-+-}|^2+2|M_{+--+}|^2+8|M_{+++-}|^2
\;.
\end{eqnarray}
\noindent Taking into account the crossing symmetry, we find
relations between amplitudes,
\begin{eqnarray}
\label{rel}
&&M_{+-+-}(\hat{s},\hat{t},\hat{u})=M_{++++}(\hat{u},\hat{t},\hat{s})
\;,
\nonumber \\
&&M_{+--+}(\hat{s},\hat{t},\hat{u})=M_{++++}(\hat{t},\hat{s},\hat{u})
= M_{++++}(\hat{t},\hat{u},\hat{s}) \;,
\nonumber \\
&&M_{+--+}(\hat{s},\hat{t},\hat{u})=M_{+-+-}(\hat{s},\hat{t},\hat{s})
\end{eqnarray}
\noindent All of SM helicity amplitudes can be found in
\cite{sm3,sm4}. Using relations $\ln(\hat{u})=\ln(-\hat{u})+i\pi, \
\ln(\hat{t})=\ln(-\hat{t})+i\pi, \ \ln(-\hat{s})=\ln(\hat{s})+
i\pi$, the helicity amplitudes corresponding to the fermion loops
can be obtained by neglecting the terms like $m_{f}^{2}/\hat{s}$,
$m_{f}^{2}/\hat{t}$ and $m_{f}^{2}/\hat{u}$
\begin{eqnarray}
\frac{1}{\alpha^{2}Q_{f}^{4}}M_{++++}^{f}(\hat{s},\hat{t},\hat{u})
=-8-8(\frac{\hat{u}-\hat{t}}{\hat{s}})\,\mathrm{Ln}({\frac{\hat{u}}{\hat{t}}})
 -4(\frac{\hat{t}^{2}+\hat{u}^{2}}{\hat{s}^{2}})
[\,\mathrm{Ln}({\frac{\hat{u}}{\hat{t}}})\,\mathrm{Ln}({\frac{\hat{u}}{\hat{t}}})+\pi^{2}]
\;,
\end{eqnarray}
\begin{eqnarray}
M_{+++-}^{f}(\hat{s},\hat{t},\hat{u})\simeq
M_{++--}^{f}(\hat{s},\hat{t},\hat{u})\simeq 8\alpha^{2}Q_{f}^{4} \;.
\end{eqnarray}
\noindent where invariant Mandelstam  variables are defined as
$\hat{s}=(p_{1}+p_{2})^{2}$, $\hat{t}=(p_{1}-p_{3})^{2}$ and
$\hat{u}=(p_{2}-p_{3})^{2}$ and $m_{f}$, $Q_{f}$ is the mass of the
fermion $f$ and its charge, respectively. The other helicity
amplitudes can be obtained by using relations in Eq.~(\ref{rel}).

It can be found the terms for $W$ loop contribution with neglecting
the $m_{W}^{2}/\hat{s}$, $m_{W}^{2}/\hat{t}$ and $m_{W}^{2}/\hat{u}$
using similar approximation,
\begin{eqnarray}
\frac{1}{\alpha^{2}}M_{++++}^{W}(\hat{s},\hat{t},\hat{u})&&=-16i\pi
[\frac{\hat{s}}{\hat{t}}\,\mathrm{Ln}(\frac{-\hat{t}}{m_{W}^{2}})
+\frac{\hat{s}}{\hat{u}}\,\mathrm{Ln}(\frac{-\hat{u}}{m_{W}^{2}})]
\nonumber \\
&&+12+12(\frac{\hat{u}-\hat{t}}{\hat{s}})\,\mathrm{Ln}(\frac{\hat{u}}{\hat{t}})
\nonumber\\
&&+16(1-\frac{3\hat{t}\hat{u}}{4\hat{s}^{2}})
[\,\mathrm{Ln}(\frac{\hat{u}}{\hat{t}})\,\mathrm{Ln}(\frac{\hat{u}}{\hat{t}})+\pi^{2}]
\nonumber \\
&&+16[\frac{\hat{s}}{\hat{t}}\,\mathrm{Ln}(\frac{\hat{s}}{m_{W}^{2}})
\,\mathrm{Ln}(\frac{-\hat{t}}{m_{W}^{2}})+
\frac{\hat{s}}{\hat{u}}\,\mathrm{Ln}(\frac{\hat{s}}{m_{W}^{2}})
\,\mathrm{Ln}(\frac{-\hat{u}}{m_{W}^{2}})\nonumber \\
&&+\frac{\hat{s}^{2}}{\hat{t}\hat{u}}\,\mathrm{Ln}(\frac{-\hat{t}}{m_{W}^{2}})
\,\mathrm{Ln}(\frac{-\hat{u}}{m_{W}^{2}})] \;,
\end{eqnarray}
\begin{eqnarray}
\frac{1}{\alpha^{2}}M_{+-+-}^{W}(\hat{s},\hat{t},\hat{u})&&=
-i\pi[12(\frac{\hat{s}-\hat{t}}{\hat{u}}) +
32(1-\frac{3\hat{t}\hat{s}}{4\hat{u}^{2}})
\,\mathrm{Ln}(\frac{\hat{s}}{-\hat{t}})\nonumber \\
&&+
16\frac{\hat{u}}{\hat{s}}\,\mathrm{Ln}(\frac{-\hat{u}}{m_{W}^{2}}) +
16\frac{\hat{u}^2}{\hat{t}\hat{s}}\,\mathrm{Ln}(\frac{-\hat{t}}{m_{W}^{2}})]
+ 12
\nonumber \\
&&+
12(\frac{\hat{s}-\hat{t}}{\hat{u}})\,\mathrm{Ln}(\frac{\hat{s}}{-\hat{t}})
+16(1-\frac{3\hat{t}\hat{s}}{4\hat{u}^{2}})
\,\mathrm{Ln}(\frac{\hat{s}}{-\hat{t}})\,\mathrm{Ln}(\frac{\hat{s}}{-\hat{t}})
\nonumber \\
&&+16[\frac{\hat{u}}{\hat{t}}\,\mathrm{Ln}(\frac{-\hat{u}}{m_{W}^{2}})
\,\mathrm{Ln}(\frac{-\hat{t}}{m_{W}^{2}})
+\frac{\hat{u}}{\hat{s}}\,\mathrm{Ln}(\frac{-\hat{u}}{m_{W}^{2}})
\,\mathrm{Ln}(\frac{\hat{s}}{m_{W}^{2}})\nonumber \\
&&+\frac{\hat{u}^{2}}{\hat{t}\hat{s}}\,\mathrm{Ln}(\frac{-\hat{t}}{m_{W}^{2}})
\,\mathrm{Ln}(\frac{\hat{s}}{m_{W}^{2}})] \;,
\end{eqnarray}
\begin{align}
M_{+-+-}^{W}(\hat{s},\hat{t},\hat{u}) &=
M_{-+-+}^{W}(\hat{s},\hat{t},\hat{u}) \;,
\nonumber \\
M_{+++-}^{W}(\hat{s},\hat{t},\hat{u}) & \simeq
M_{++--}^{W}(\hat{s},\hat{t},\hat{u})\simeq -12\alpha^{2} \;.
\end{align}

In case of $m_{W}^{2} \ll \hat{s}$, $W$ loop helicity amplitudes
(especially, their imaginary parts) become dominant. In $m_{W}^{2}
\gg \hat{s}$ energy region fermion loop contributions are much
bigger than the $W$ loop contributions. The contribution of the top
quark in all energy region is not taken into account since it is
very small compared to other fermions and $W$ loop contributions
\cite{Atag:2010}.

%%%%%%%%%%%%%%%%%%%%%%%%%%%%%%%%%%%%%%%%%%%%%%%%%%%%
\section{Photon-Photon interactions at the LHC} %
%%%%%%%%%%%%%%%%%%%%%%%%%%%%%%%%%%%%%%%%%%%%%%%%%%%%
\label{subsec:gamma-gamma}

As it was mentioned above, it is possible to examine the
photon-photon interaction with using forward detectors at the LHC.
After elastic photon emission with small angels and low transverse
momentum, the protons deviate slightly from their paths along the
beam pipe and are probed in the forward detectors without being
detected by main detectors. This deviation is related to $\xi$.
Emitted photons which are called almost-real photons have very low
virtualities. Therefore, these photons can be considered
on-mass-shell. In this case, the process $pp \to p \gamma \gamma p
\to p' \gamma \gamma p'$ occurs and the final state $X$ is measured
by the central detector. The value of $\xi$ can be determined by
using forward detectors. Hence, the center of mass energy of the
$\gamma\gamma$ collision can be known. It is given as
$W=2E\sqrt{\xi_1 \xi_2}$, where $E$ is the energy of the incoming
protons with the mass $m_{p}$. The photon-photon interaction in the
hadron collision can be studied with the equivalent photon
approximation (EPA) \cite{budnev,baur}. The EPA includes a spectrum
that depend on the photon energy ($E_\gamma=\xi E$) and photon
virtuality ($Q^2 = -q^2$)
\begin{eqnarray}
\frac{dN_\gamma}{dE_{\gamma}dQ^{2}}=\frac{\alpha}{\pi}\frac{1}{E_{\gamma}Q^{2}}
\left[ (1-\frac{E_{\gamma}}{E}) (1 -
\frac{Q^{2}_{\min}}{Q^{2}})F_{E}(Q^2) +
\frac{E^{2}_{\gamma}}{2E^{2}}F_{M}(Q^2) \right] . \label{phs}
\end{eqnarray}

\begin{figure}[htb]
\begin{center}
\includegraphics[scale=0.60]{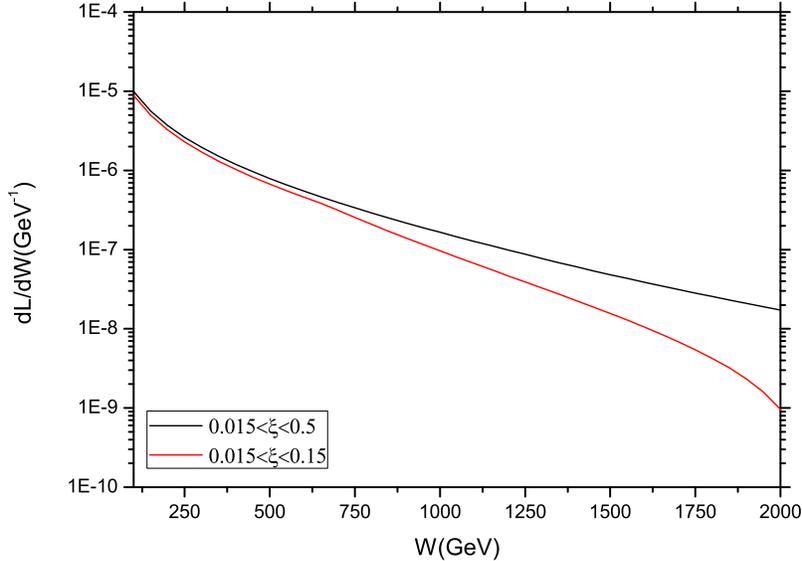}
\caption{Effective $\gamma\gamma$ luminosity as a function of the
invariant mass of the two photon system. Figure shows the effective
luminosity for two forward detector acceptances, $0.015 < \xi < 0.5$
and $0.015 < \xi < 0.15$.}
\label{fig:lum}
\end{center}
\end{figure}

\noindent The minimal photon virtuality $Q^{2}_{\min}$, as well as
electric ($F_{E}$) and magnetic ($F_{M}$) form factors of the proton
in above equation are defined in Ref.~\cite{Atag:2010}. From this
perspective, the resulting luminosity spectrum
$dL^{\gamma\gamma}/dW$ is obtained as
\begin{eqnarray}
\label{efflum}
\frac{dL^{\gamma\gamma}}{dW}=\int_{Q^{2}_{1,\min}}^{Q^{2}_{\max}}
{dQ^{2}_{1}}\int_{Q^{2}_{2,\min}}^{Q^{2}_{\max}}{dQ^{2}_{2}}
\int_{y_{\min}}^{y_{\max}} {dy \frac{W}{2y} f_{1}(\frac{W^{2}}{4y},
Q^{2}_{1}) f_{2}(y,Q^{2}_{2})} \;,
\end{eqnarray}

\noindent with $y_{\min}=\mbox{max}(W^{2}/(4\xi_{\max}E),\,
\xi_{\min}E)$, $y_{\max}=\xi_{\max}E$, $f = dN/(dE_{\gamma}dQ^{2})$,
$Q_{\max}^2=2$ GeV. The contribution of more than this $Q_{\max}^2$
value is negligible to the integral. In Fig.~\ref{fig:lum}, we show
the effective $\gamma\gamma$ luminosity as a function of $W$ for the
detector acceptances $0.015<\xi<0.5$ and $0.015<\xi<0.15$.
%%%%%%%%%%%%
% Figure 2 %
%%%%%%%%%%%%
Using the Eq.~\ref{efflum} the total cross section for the $pp \to p
\gamma \gamma p \to p' \gamma\gamma p' $ can be given as follows
\begin{eqnarray} \label{completeprocess}
d\sigma=\int{\frac{dL^{\gamma\gamma}}{dW}
\,d\hat{{\sigma}}_{\gamma\gamma \to \gamma\gamma}(W)\,\,dW} \;,
\end{eqnarray}
where  $d\hat {{\sigma}}_{\gamma\gamma \to \gamma\gamma}(W)$ is the
cross section of the subprocess $\gamma\gamma \to \gamma\gamma$.

%%%%%%%%%%%%%%%%%%%%%%%%%%%%%%%%%%%%%%%%%%%%%%%%%%%%%%%%%%%%%%
\section{Randall-Sundrum--like model with a small curvature} %
%%%%%%%%%%%%%%%%%%%%%%%%%%%%%%%%%%%%%%%%%%%%%%%%%%%%%%%%%%%%%%
\label{sec:RSSC}

One of promising possibilities to go beyond the SM is to consider a
scenario with extra spatial dimensions (EDs). A framework with EDs
is motivated by the (super)string theory \cite{Polchinski:98}. One
of the main goals of such theories is to explain the hierarchy
relation between electromagnetic and Planck scales. In the model
proposed by Arkani-Hamed, Dimopolous, Dvali and Antoniadis
\cite{Arkani-Hamed:1998}--\cite{Hamed2:1998}, called ADD, the
hierarchy relation looks like
\begin{equation}\label{ADD_hierarchy}
\bar{M}_{\mathrm{Pl}}^2 = V_d M_D^{d+2} \;,
\end{equation}
where $V_d = (2\pi R_c)^d$ is the volume of the compactified extra
dimensions with the size $R_c$, $\bar{M}_{\mathrm{Pl}} =
M_{\mathrm{Pl}}/\sqrt{8\pi}$ is the reduced Planck mass, and $M_D$
is the fundamental gravity scale in $D=4+d$ dimensions. The masses
of the Kaluza-KLein (KK) gravitons in the ADD model are
\begin{equation}\label{ADD_masses}
m_n = \frac{n}{R_c}, \quad n = \sqrt{n_1^2 + n_2^2 + \cdots n_d^2}
\;,
\end{equation}
where $n_i = 0, 1, \ldots$ ($i=1,2, \ldots d$). Thus, in the
scenario with large EDs the mass splitting $\Delta m_{KK} = 1/R_c$
is very small.

However, this solution of the hierarchy problems in the ADD model
cannot be considered satisfactory, since formula
\eqref{ADD_hierarchy} explains a large value of the Planck mass by
introducing new large scale, the volume of EDs. To overcome this
shortcoming, the model with one warped ED and two branes, known as
RS1, was proposed by Randall and Sundrum \cite{Randall:1999}.

The RS1 model is described by the following background warped metric
\begin{equation}\label{RS_metric}
\quad ds^2 = e^{-2 \sigma (y)} \, \eta_{\mu \nu} \, dx^{\mu} \,
dx^{\nu} - dy^2 \;,
\end{equation}

\noindent where $\eta_{\mu\nu}$ is the Minkowski tensor with the
signature $(+,-,-,-)$, and $y$ is an extra coordinate. The
periodicity condition $y=y + 2\pi r_c$ is imposed, and the points
$(x_\mu,y)$ and $(x_\mu,-y)$ are identified. As a result, we have a
model of gravity in a slice of the AdS$_5$ space-time compactified
to the orbifold $S^1\!/Z_2$. The orbifold has two fixed points,
$y=0$ and $y=\pi r_c$. Two branes are located at these points
(called Planck and TeV brane). All the SM fields are assumed to live
on the TeV brane.

The classical action of the RS1 model is \cite{Randall:1999}
\begin{align}\label{action}
S &= \int \!\! d^4x \!\! \int_{-\pi r_c}^{\pi r_c} \!\! dy \,
\sqrt{G} \, (2 \bar{M}_5^3 \mathcal{R}
- \Lambda) \nonumber \\
&+ \int \!\! d^4x \sqrt{|g^{(1)}|} \, (\mathcal{L}_1 - \Lambda_1) +
\int \!\! d^4x \sqrt{|g^{(2)}|} \, (\mathcal{L}_2 - \Lambda_2) \;,
\end{align}

\noindent where $G_{MN}(x,y)$ is the 5-dimensional metric, $M,N =
0,1,2,3,4$. The quantities $g^{(1)}_{\mu\nu}(x) = G_{\mu\nu}(x,
y=0)$, $g^{(2)}_{\mu\nu}(x) = G_{\mu\nu}(x, y=\pi r_c)$, where $\mu
= 0,1,2,3$, are induced metrics on the branes, $\mathcal{L}_1$,
$\mathcal{L}_2$ are brane Lagrangians, $G = \det(G_{MN})$, and
$g^{(i)} = \det(g^{(i)}_{\mu\nu})$ ($i=1,2$). $\bar{M}_5$ is the
reduced 5-dimensional Planck scale, $M_5/(2\pi)^{1/3}$, $M_5$ being
the fundamental gravity scale in five dimensions. $\Lambda$ is a
5-dimensional cosmological constant, while $\Lambda_{1,2}$ are brane
tensions.

The warp function $\sigma(y)$ in eq.~\eqref{RS_metric} obeys
Einstein-Hilbert's equations. For the first time, it was derived in
\cite{Randall:1999} to be $\sigma_{\mathrm{RS}}(y) = \kappa |y|$,
where $\kappa$ is a parameter with a dimension of mass. It defines
the curvature of the 5-dimensional space-time, $\mathcal{R} = - 20
\kappa^2$.

The hierarchy relation in the RS1 model is of the form
\cite{Randall:1999}
\begin{equation}\label{RS1_hierarchy}
\bar{M}_{\mathrm{Pl}}^2 = \frac{\bar{M}_5^3}{\kappa} \left[ 1 -
e^{-2\pi \kappa r_c} \right] \Big|_{\kappa \pi r_c \gg 1} =
\frac{\bar{M}_5^3}{\kappa} \;.
\end{equation}
In order this relation to be satisfied, one has to put $\bar{M}_5
\sim \kappa \sim \bar{M}_{\mathrm{Pl}}$. It was shown that $0.01 <
\kappa/\bar{M}_5 < 0.1$ \cite{Davoudiasl:2001}. As a result,
experimental signature of the RS1 model is a series of heavy
resonances, with masses defined by the formula
\begin{equation}\label{RS1_graviton_masses}
m_n = x_n \kappa \,e^{-\pi \kappa r_c} \;, \quad n=1,2, \ldots \;,
\end{equation}
where $x_n$ are zeros of the Bessel function $J_1(x)$.

In \cite{Kisselev:2016} a general solution for $\sigma(y)$ was
derived. It looks like
\begin{equation}\label{sigma}
\sigma(y) = \frac{\kappa r_c}{2} \left[ \left| \mathrm{Arccos}
\left(\cos \frac{y}{r_c} \right) \right| - \left| \pi -
\mathrm{Arccos} \left(\cos \frac{y}{r_c} \right)\right| \right] +
\frac{\pi \,|\kappa| r_c }{2} - C \;,
\end{equation}
where $\mathrm{Arccos(z)}$ is a principal value of the multivalued
inverse trigonometric function $\arccos(z)$, and $C$ is
$y$-independent arbitrary parameter. By taking $C=0$ in
\eqref{sigma}, we reproduce the RS1 model, while putting $C = \pi
\kappa r_c$, we come to the Randall-Sundrum-like scenario with a
small curvature of the space-time (RSSC model, for details, see
\cite{Giudice:2005,Kisselev:2006}). It was applied for exploring a
number of processes at the LHC \cite{Kisselev:2008,Kisselev:2013}.

Let us see what the main features of the RSSC model are in
comparison with the features of the RS1 model. The interactions of
the KK gravitons $h_{\mu\nu}^{(n)}$ with the SM fields on the TeV
brane are given by the following effective Lagrangian density
\begin{equation}\label{Lagrangian}
\mathcal{L}_{\mathrm{int}} = - \frac{1}{\bar{M}_{\mathrm{Pl}}} \,
h_{\mu\nu}^{(0)}(x) \, T_{\alpha\beta}(x) \, \eta^{\mu\alpha}
\eta^{\nu\beta} - \frac{1}{\Lambda_\pi} \sum_{n=1}^{\infty}
h_{\mu\nu}^{(n)}(x) \, T_{\alpha\beta}(x) \, \eta^{\mu\alpha}
\eta^{\nu\beta} \;,
\end{equation}
where $T^{\mu \nu}(x)$ is the energy-momentum tensor of the SM
fields (recall that all SM fields are confined on the TeV brane).
The coupling constant is given as
\begin{equation}\label{Lambda}
\Lambda_\pi = \left( \frac{\bar{M}_5^3}{\kappa} \right)^{\!1/2} .
\end{equation}
In the RSSC model the hierarchy relation takes the form
\begin{equation}\label{hierarchy}
\bar{M}_{\mathrm{Pl}}^2 = \frac{\bar{M}_5^3}{\kappa} \left[ e^{2\pi
\kappa r_c} - 1 \right] \Big|_{\kappa \pi r_c \gg 1} =
\frac{\bar{M}_5^3}{\kappa} \,e^{2\pi \kappa r_c} \;.
\end{equation}
This relation should be compared with eq.~\eqref{RS1_hierarchy}. The
masses of the KK gravitons are equal to
\cite{Giudice:2005,Kisselev:2005}
\begin{equation}\label{graviton_masses}
m_n = x_n \kappa \;, \quad n=1,2, \ldots \;.
\end{equation}
If we take $\kappa \ll \bar{M}_5 \sim 1$ TeV, we obtain an almost
continuous graviton mass spectrum, which is similar to the spectrum
of the ADD model \eqref{ADD_masses}, since $\Delta m_{KK} \simeq \pi
\kappa$. Let us recall that in the RS1 model the KK gravitons are
heavy resonances with masses above few TeV.

Since in the RSSC scenario the warp factor $e^{-2 \sigma (y)}$ is
equal to unity on the TeV brane ($y=\pi r_c$), the coordinates on
this brane are Galilean, and the four-dimensional graviton field
$h_{\mu\nu}^{(n)}(x)$ couples to energy-momentum of the ordinary
matter $T_{\mu\nu}(x)$ in the usual way \cite{Rubakov:2001}. The
Einstein tensor $R_{\mu\nu} - (1/2)R g_{\mu\nu}$ is invariant under
transformation $\sigma(y) \rightarrow \sigma(y) - C$. As for
energy-momentum tensor, it is invariant only for massless fields.
The invariance of the gravity action under such transformation needs
rescaling of the graviton fields and their masses: $h^{(n)}_{\mu\nu}
= e^{-C} h'^{(n)}_{\mu\nu}$, $m_n = e^{-C} m'_n$. We see that the
theory of massive KK gravitons is not scale-invariant. Only its zero
mass sector (standard gravity) remains unchanged. More details can
be found in \cite{Kisselev:2016}.

Sometimes it is convenient to work with a conformally flat metric by
introducing the coordinate $z = \kappa^{-1} e^{\sigma(y)}$
\cite{Chamblin:2000}. Then the reduced Planck scale reads
$\bar{M}_{\mathrm{Pl}}^2 = (\bar{M}_5/\kappa)^3 (z_1^{-2} -
z_2^{-2})$, and the KK graviton mass is given as $m_m=x_n z_2^{-1}$,
where $z_1(z_2)$ is the conformal coordinate of the Planck(TeV)
brane. In the RS1 model $z_1\kappa = 1$, and $z_2\kappa =
e^{\pi\kappa r_c}$ (see, correspondingly,
eqs.~\eqref{RS1_hierarchy}, \eqref{RS1_graviton_masses}). On the
contrary, in the RSSC model $z_1\kappa = e^{-\pi\kappa r_c}$, and
$z_2\kappa = 1$ (see eqs.~\eqref{hierarchy},
\eqref{graviton_masses}). Note that the exponential hierarchy
between the branes is the same in both models, $z_2/z_1 =
e^{\pi\kappa r_c}$.

Now let us consider the $s$-channel KK graviton exchange
contribution to the matrix element of the subprocess $\gamma \gamma
\rightarrow \gamma \gamma$ with the invariant energy
$\sqrt{\hat{s}}$. It is defined by the formula
\begin{align}\label{KK_matrix_element}
M_{KK} &= \frac{1}{2\Lambda_\pi^2} \sum_{n=1}^\infty
e_{\gamma}(p_{1})e_{\delta}(p_{2})
\,\Gamma^{\mu\nu\gamma\delta}(p_1, p_2)
\,\frac{B_{\mu\nu\alpha\beta}}{\hat{s} - m^{2}_{n} + i\Gamma_n} \nonumber \\
&\times \Gamma^{\alpha\beta\rho\sigma}(k_1, k_2)
\,e_{\rho}(k_{1})e_{\sigma}(k_{2})] \;,
\end{align}

\noindent where $k_i$, $p_i$ ($i=1,2$) are momenta of incoming and
outgoing photons, while $e_{\mu}(k_{i})$, $e_{\mu}(p_{i})$  are
their polarization vectors. $\Gamma^{\alpha\beta\rho\sigma}$ is a
$h^{(n)} \gamma\gamma$ vertex function, $B_{\mu\nu\alpha\beta}$ is a
tensor part of the graviton propagator. Explicit forms of the
tensors $\Gamma^{\alpha\beta\rho\sigma}$ and $B_{\mu\nu\alpha\beta}$
can be found in Ref.~\cite{Inan-Kisselev}. The coherent sum in
\eqref{KK_matrix_element} is over KK modes. The total width of the
graviton with the KK number $n$ and mass $m_n$ is given by $\Gamma_n
= 0.09 \,m_n^3/\Lambda_\pi^2$ \cite{Kisselev:2005_2}.

Let us concentrate on the scalar part of the sum
\eqref{KK_matrix_element} which is universal for all types of
processes mediated by the s-channel exchanges of the KK gravitons.
It is of the form:
\begin{equation}\label{S_s}
\mathcal{S}(s) = \frac{1}{\Lambda_\pi^2} \sum_{n=1}^\infty
\frac{1}{s - m^{2}_{n} + i\Gamma_n} \;.
\end{equation}
This sum has been calculated in \cite{Kisselev:2006}
\begin{equation}\label{S_s_exp}
\mathcal{S}(s) = - \frac{1}{4\bar{M}_5^3 \sqrt{s}} \; \frac{\sin
(2A) + i \sinh (2\varepsilon)}{\cos^2 \!A + \sinh^2 \! \varepsilon }
\;.
\end{equation}
where
\begin{equation}\label{A_epsilon}
A = \frac{\sqrt{s}}{\kappa} \;, \qquad \varepsilon  = 0.045 \left(
\frac{\sqrt{s}}{\bar{M}_5} \right)^{\!\!3} .
\end{equation}

As for the contribution from the $t$-channel graviton exchanges,
$\mathcal{S}(\hat{t})$, is was shown in \cite{Kisselev:2006} that
the function $\mathcal{S}(t)$ is pure real for $t<0$, $\bar{M}_5 \gg
\kappa$
\begin{equation}\label{S_t}
\mathcal{S}(t) = - \frac{1}{2\bar{M}_5^3 \sqrt{-t}} \;.
\end{equation}
Analogously, we have for the $u$-channel graviton exchanges
\begin{equation}\label{S_u}
\mathcal{S}(u) = - \frac{1}{2\bar{M}_5^3 \sqrt{-u}} \;.
\end{equation}
Let us underline that a magnitude of the matrix element is defined
by the fundamental gravity scale $\bar{M}_5$, not by the coupling
constant $\Lambda_\pi$ \eqref{Lambda}.

The virtual KK graviton exchanges should lead to deviations from the
SM predictions both in a magnitude of the cross sections and angular
distribution of the final photons because of the spin-2 nature of
the gravitons. For the ADD model, the pure KK graviton contribution
to the matrix element of the subprocess $\gamma \gamma \rightarrow
\gamma \gamma$ was calculated in \cite{Atag:2010}. Its
generalization for the RSSC model looks like
\begin{align}\label{M2_KK}
|M_{KK}|^2 &= \frac{1}{8} \big\{ |\mathcal{S}(\hat{s})|^2 (\hat{t}^4
+ \hat{u}^4) + |\mathcal{S}(\hat{t})|^2 (\hat{s}^4 + \hat{u}^4) +
|\mathcal{S}(\hat{u})|^2 (\hat{s}^4 + \hat{t}^4) \nonumber\\
&+ [ \mathcal{S}(\hat{s})^\star \mathcal{S}(\hat{t}) +
\mathcal{S}(\hat{s}) \mathcal{S}^\star(\hat{t}) ] \hat{u}^4 + [
\mathcal{S}(\hat{s})^\star \mathcal{S}(\hat{u}) +
\mathcal{S}(\hat{s}) \mathcal{S}^\star(\hat{u}) ] \hat{t}^4 \nonumber\\
&+ [ \mathcal{S}^\star(\hat{t}) \mathcal{S}(\hat{u}) +
\mathcal{S}(\hat{t}) \mathcal{S}^\star(\hat{u})] \big\} \;,
\end{align}
where $\hat{s}$, $\hat{t}$, $\hat{u}$ are Mandelstam variables of
the subprocess $\gamma \gamma \rightarrow \gamma \gamma$, and the
functions $\mathcal{S}(s), \mathcal{S}(t)$, $\mathcal{S}(u)$ are
defined above.

%%%%%%%%%%%%%%%%%%%%%%%%%%%%%%%%%%%%%%%%%%%%%%%%%%%%%%%%%%%%%%%%%%%%%

%%%%%%%%%%%%%%%%%%%%%%%%%%%%%%
\section{Numerical analysis} %
%%%%%%%%%%%%%%%%%%%%%%%%%%%%%%
\label{sec:num_analysis}

As it was mentioned above, in the RSSC model the KK graviton
spectrum is similar to that in the ADD model. That is why, in
contrast to the original RS1 model, an account of effects from EDs
in the RSSC model leads to deviations from the SM in magnitudes both
of differential cross sections and total cross sections for the
photon-induced process $pp \rightarrow p \gamma\gamma p \rightarrow
p' \gamma\gamma p'$ at the LHC. This process goes via electroweak
subprocess $\gamma\gamma \rightarrow \gamma\gamma$.

Our main goal is to calculate these deviations as a function of the
parameters of the RSSC model. It will enable us to set the 95\% C.L.
search limits for the reduced 5-dimensional Planck scale
$\bar{M}_5$. Let us underline that this limits don't depend (up to
small power corrections $\sim \kappa/\bar{M}_5$) on a value of the
second parameter of the model $\kappa$. It is an interesting feature
of the RS-like scenario with the small curvature.

Since we impose the cut $W > 200$ GeV on the diphoton invariant
mass, we can neglect the QCD loop contributions (see
Section~\ref{sec:intr}). Below, to estimate the LHC search limit, we
will take the cut $p_t > 300(500)$ GeV, where $p_t$ is the final
photon transverse momentum. Note that $W \geqslant 2p_t$ due to
energy conservation. Thus, the condition $W
> 200$ GeV will be automatically satisfied.

We also impose the cut $|\eta_{pp}^{i}| < 2.5$ on the rapidities of
the final photons $\eta_{pp}^{i}$ ($i=1,2$) in the c.m.s of the
\emph{colliding protons}. It is equivalent to the inequality
\begin{equation}\label{eta_cut}
\eta_{\gamma\gamma} + |\eta_X| < 2.5 \;,
\end{equation}
where
\begin{equation}\label{photon_rapididities}
\eta_{\gamma\gamma} = \ln \!\frac{W + \sqrt{W^2 - 4p_t^2}}{2p_t}
\end{equation}
is the rapiditidy of the final photons in the c.m.s of \emph{two
photons}, and
\begin{equation}\label{eta_X}
\eta_X = \frac{1}{2} \ln \frac{\xi_1}{\xi_2}
\end{equation}
is the rapidity of the diphoton system in the in the c.m.s of the
incoming protons.

\begin{figure}[htb]
\begin{center}
\includegraphics[scale=0.5]{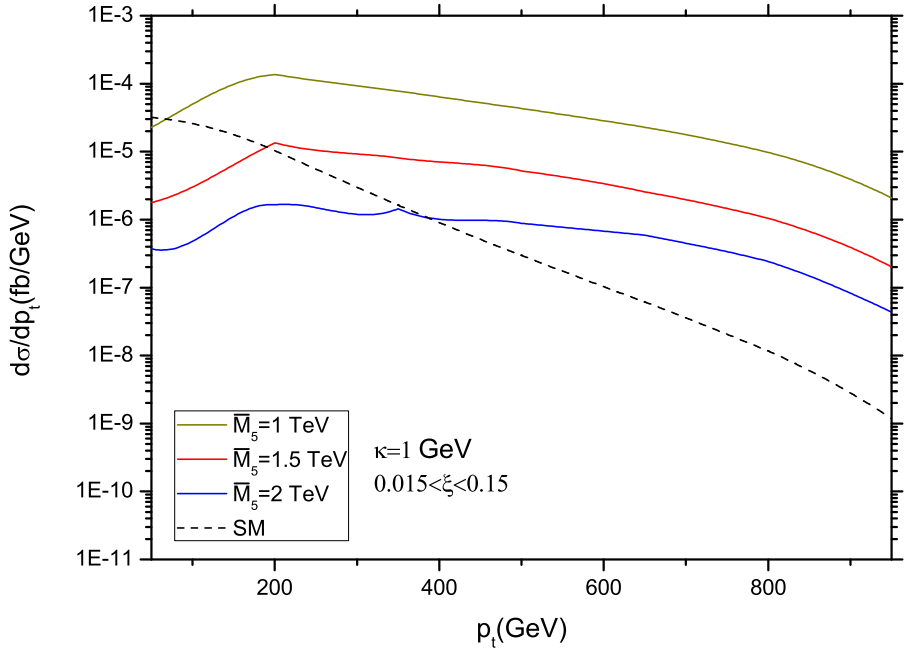}
\caption{The differential cross section for the process $pp
\rightarrow p \gamma\gamma p$ as a function of the transverse
momenta of the final photons for $\kappa = 1$ GeV and for the
acceptance region $0.015 < \xi < 0.15$. The cut on the photon
rapidities, $|\eta| < 2.5$, is imposed. Here and below the dotted
line denotes the SM contribution.} \label{fig:ptd15015_f}
\end{center}
\end{figure}

\begin{figure}[htb]
\begin{center}
\includegraphics[scale=0.5]{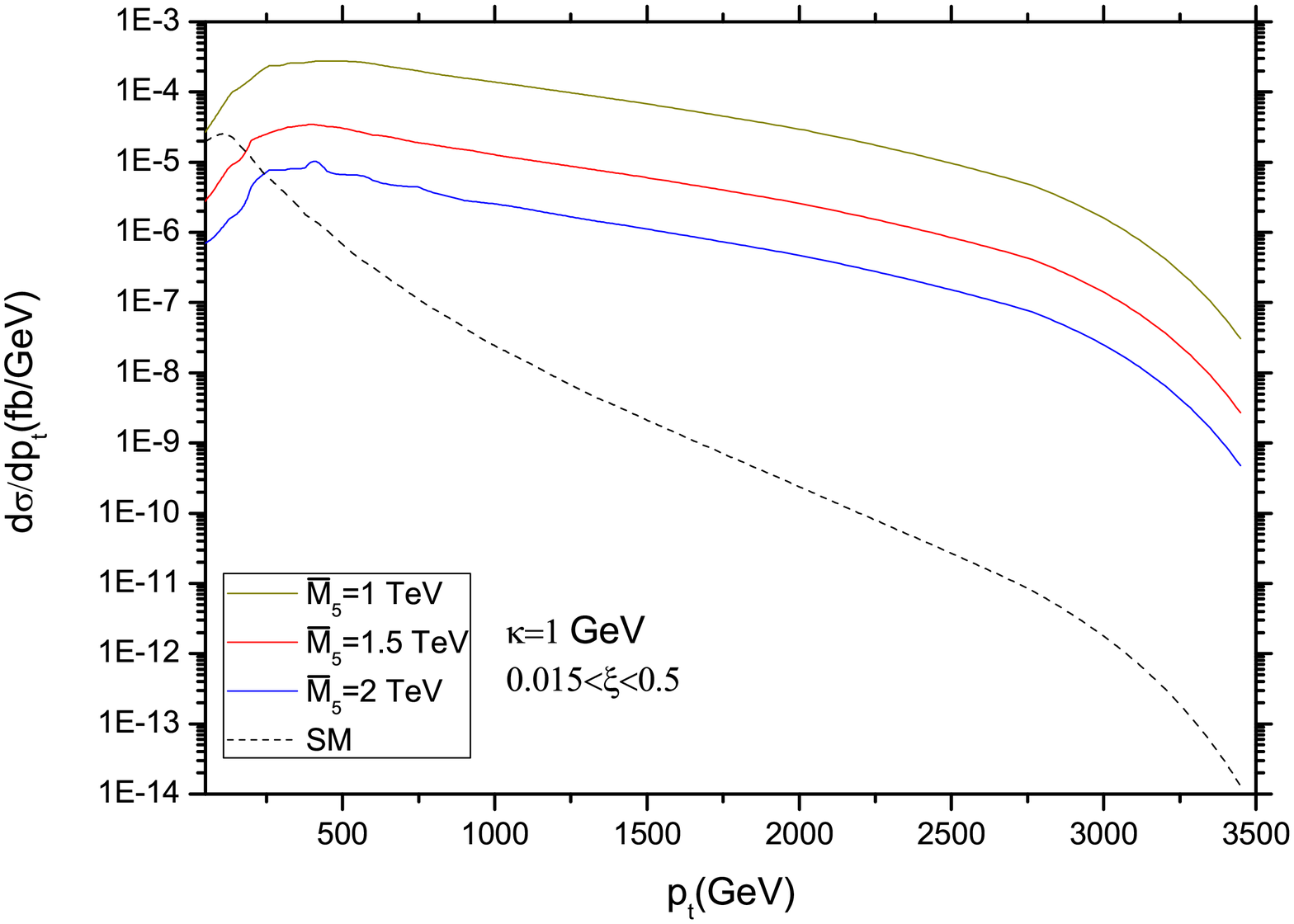}
\caption{The same as in Fig.~\ref{fig:ptd15015_f}, but for the
acceptance region $0.015 < \xi < 0.5$.} \label{fig:ptd05015_f}
\end{center}
\end{figure}

The results of our calculations of the differential cross sections
$d\sigma/dp_t$ with the cuts mentioned above as a function of the
photon transverse momenta are presented in
Figs.~\ref{fig:ptd15015_f} and \ref{fig:ptd05015_f} for three values
of $\bar{M}_5$.
%
%%%%%%%%%%%%
% Figure 3 %
%%%%%%%%%%%%
%
%%%%%%%%%%%%
% Figure 4 %
%%%%%%%%%%%%
\noindent Our calculations have shown that the differential cross section does
not practically depend on the curvature parameter $\kappa$. The same
is true for the dimuon production in photon-induced events at the
LHC \cite{Inan-Kisselev}. One can see that $d\sigma/dp_t$ exceeds
the SM cross section $d\sigma_{\mathrm{SM}}/dp_t$ for $p_t > 300$
GeV, if $0.015 < \xi < 0.15$, and for $p_t > 500$ GeV, if $0.015 <
\xi < 0.5$. Moreover, the difference between $d\sigma/dp_t$ and
$d\sigma_{\mathrm{SM}}/dp_t$ increases as $p_t$ grows. The effect is
more pronounced for smaller values of $\bar{M}_5$. The maximum of
$d\sigma/dp_t$ around $p_t \simeq 200$ GeV (500 GeV) for the
acceptance region $0.015 < \xi < 0.15$ ($0.015 < \xi < 0.5$) is a
result of the integration in variable $W$, whose lower limit depend
on $p_t$, as well as due to the $p_t$-dependence of the rapidity cut
\eqref{eta_cut}--\eqref{photon_rapididities}.

\begin{figure}[htb]
\begin{center}
\includegraphics[scale=0.5]{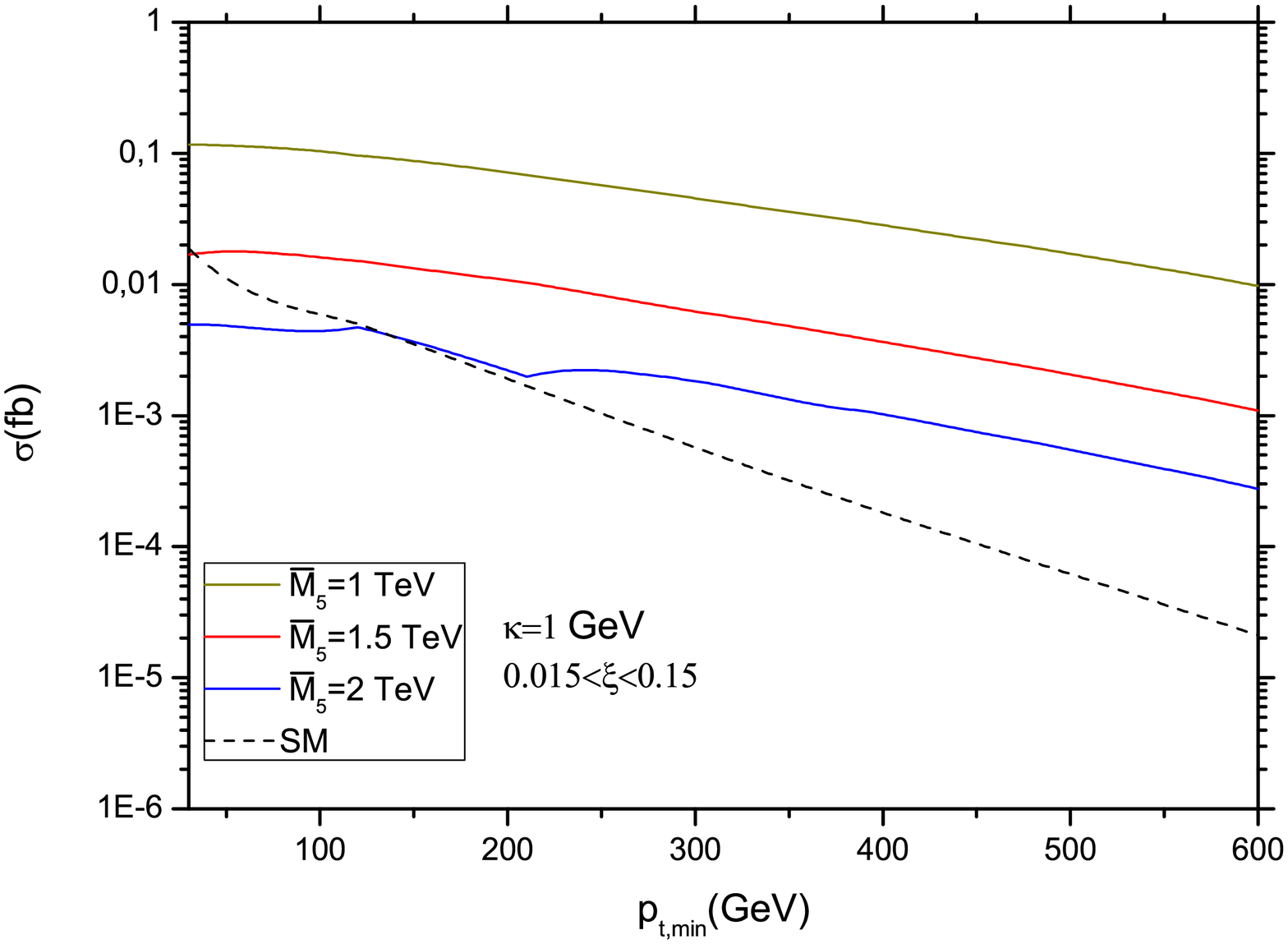}
\caption{The total cross section for the process $pp \rightarrow p
\gamma\gamma p$ as a function of the minimal transverse momenta of
the final photons $p_{t,\min}$  for the acceptance region $0.015 <
\xi < 0.15$ for different values of $\bar{M}_5$.}
\label{fig:ptcut15015_f}
\end{center}
\end{figure}

\begin{figure}[htb]
\begin{center}
\includegraphics[scale=0.5]{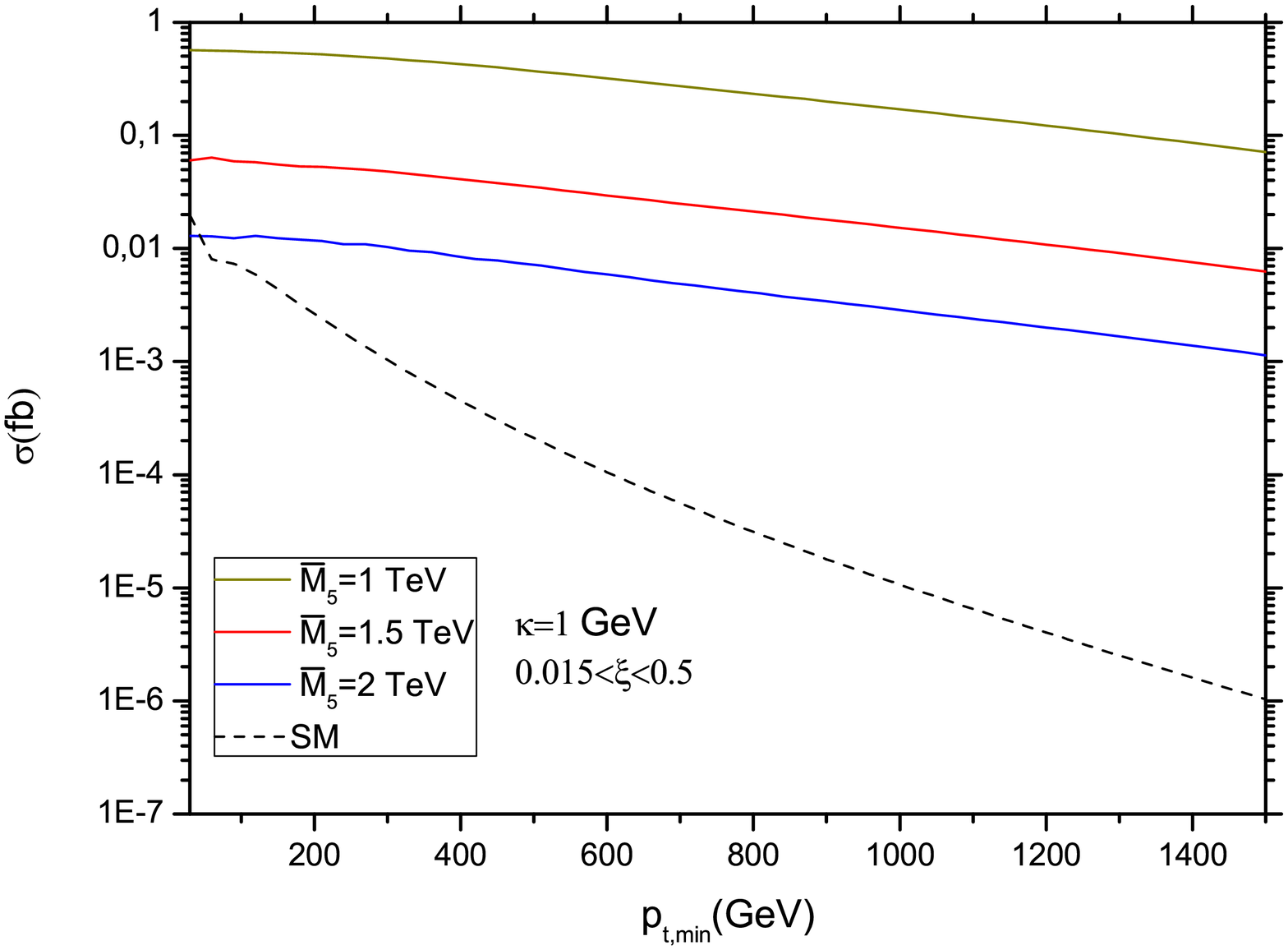}
\caption{The same as in Fig.~\ref{fig:ptcut15015_f}, but for the
acceptance region $0.015 < \xi < 0.5$.}
\label{fig:ptcut05015_f}
\end{center}
\end{figure}

The total cross section $\sigma(p_t > p_{t,\min})$ for two
acceptance regions is shown in Figs.~\ref{fig:ptcut15015_f} and
\ref{fig:ptcut05015_f} as a function of the minimal transverse
momenta of the final photons $p_{t,\min}$.
%
%%%%%%%%%%%%
% Figure 5 %
%%%%%%%%%%%%
%
%%%%%%%%%%%%
% Figure 6 %
%%%%%%%%%%%%
In both figures, the comparison with the pure SM predictions is
given. For both acceptance regions, a deviation of $\sigma(p_t >
p_{t,\min})$ from the SM cross section $\sigma_{\mathrm{SM}}(p_t >
p_{t,\min})$ gets higher as $p_{t,\min}$ grows. The effect is more
significant for $0.015 < \xi < 0.5$.

Having calculations of the total cross sections in hand, we are able
to obtain the limits on $\bar{M}_5$ for two acceptance regions,
$0.015<\xi<0.15$ and $0.015<\xi<0.5$, for $p_t > 300$ GeV and $p_t >
500$ GeV, respectively. In sensitivity analysis, we use the
likelihood method from \cite{fic3}. We assume that observed events
follow a Poisson distribution. Then the statistics together with the
prediction of the event rate leads to to the following likelihood
function
\begin{equation}\label{Likelihood}
\mathfrak{L}(\sigma) = \mathrm{Pr(n|b + \sigma L)} \;.
\end{equation}
Here $n$ is the number of the observed events, $b$ is the expected
number of background (SM) events, $\sigma$ is the total cross
section, and $L$ is the integrated luminosity. One can estimate from
Figs.~\ref{fig:ptcut15015_f} and \ref{fig:ptcut05015_f} that for the
maximum luminosity value of $L= 300$ fb$^{-1}$ ($L= 3000$ fb$^{-1}$)
when $p_t > 300$ GeV ($p_t > 500$ GeV), the expected number of the
SM events is less than 0.5. Thus, we can assume that no events is
observed, and put $b = 0$. Then the LHC exclusion region for the
credibility $1 - \alpha$ is given by the formula \cite{fic3}
\begin{equation}\label{credibility_def}
\sigma_\alpha = - \frac{1}{L} \ln (\sigma) \;.
\end{equation}
For the 95\% C.L., which corresponds to $\alpha = 0.05$, we get from
\eqref{credibility_def}
\begin{equation}\label{credibility_final}
\sigma_{0.05} \simeq \frac{3}{L} \;.
\end{equation}

\begin{figure}[htb]
\begin{center}
\includegraphics[scale=0.9]{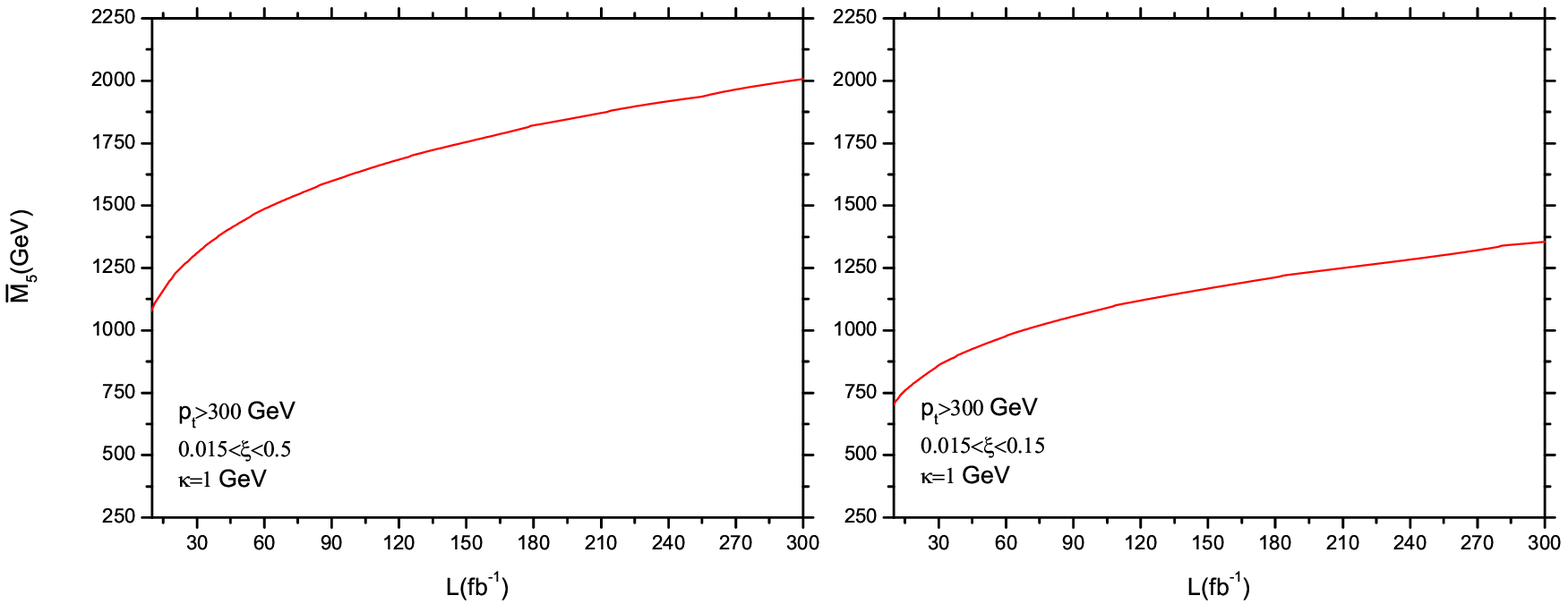}
\caption{The 95\% C.L. search limits for the reduced 5-dimensional
gravity scale $\bar{M}_5$ as a function of the integrated LHC
luminosity with $p_t > 300$ GeV. The rapidity cut of $2.5$ on the
photon rapidities are imposed. Left panel: $0.015 < \xi < 0.15$.
Right panel: $0.015 < \xi < 0.5$.}
\label{fig:CL_pt300f}
\end{center}
\end{figure}

\begin{figure}[htb]
\begin{center}
\includegraphics[scale=0.9]{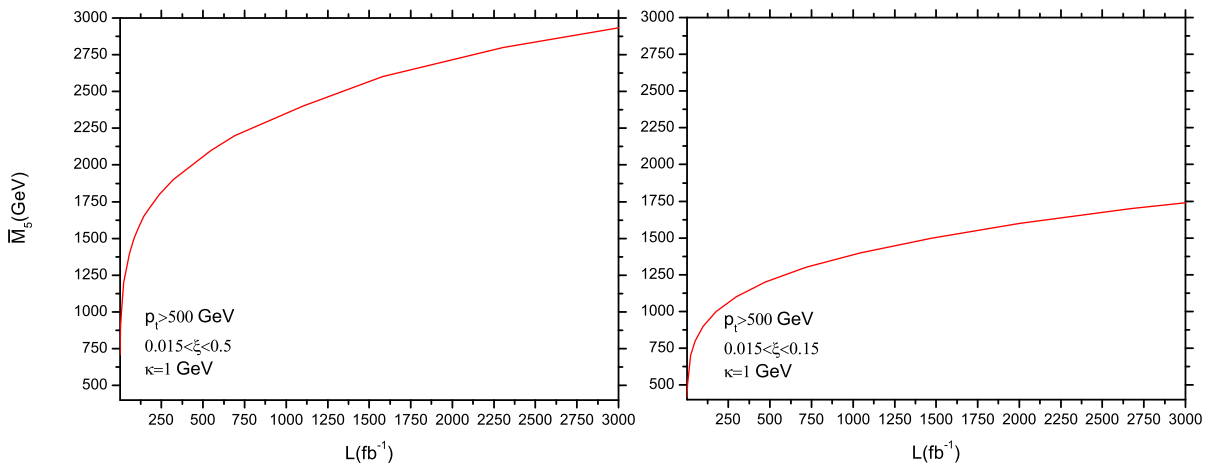}
\caption{The same as in Fig.~\ref{fig:CL_pt300f}, but for the
acceptance region $p_t > 500$ GeV.}
\label{fig:CL_pt500f}
\end{center}
\end{figure}

First, let us consider the acceptance region $0.015 < \xi < 0.15$
and impose the cut $p_t > 300$ GeV. Using
eq.~\eqref{credibility_def}, we have found the 95\% C.L. search
limits for the reduced 5-dimensional gravity scale $\bar{M}_5$ as a
function of the integrated LHC luminosity, see
Fig.~\ref{fig:CL_pt300f}. The analogous results for the cut $p_t >
500$ GeV are presented in Fig.~\ref{fig:CL_pt500f}. As one can see,
for the integrated luminosity $L= 300$ fb$^{-1}$, the sensitivity
bounds on $\bar{M}_5$ are 2.01 TeV and 1.37 TeV, for the acceptance
region $0.015 < \xi < 0.5$ and $0.015 < \xi < 0.15$, respectively.
For $L= 3000$ fb$^{-1}$ the sensitivity bounds on $\bar{M}_5$ are
equal to 2.93 TeV and 1.74 TeV. Let us underline that these bound
don't depend on the parameter $\kappa$, provided $\kappa \ll
\bar{M}_5$, what is satisfied in our analysis.

%%%%%%%%%%%%
% Figure 7 %
%%%%%%%%%%%%
%
%%%%%%%%%%%%
% Figure 8 %
%%%%%%%%%%%%

Our bounds on the 5-dimensional gravity scale $\bar{M}_5$ are rather
low in comparison with the experimental bounds on $D$-dimensional
scale $M_D$ in the ADD model (see, for instance,). In this regard,
we must emphasize that the LHC bounds on $M_D$ cannot be directly
applied to the gravity scale $\bar{M}_5$ in the RSSC model. As was
mentioned above (for details, see \cite{Kisselev:2006}), this model
cannot be regarded as a small distortion of the ADD model even for
very small values of the curvature $\kappa$. Moreover, in the ADD
model the number of EDs should be $d \geqslant 2$, while in the RSSC
model we deal with one ED. As for the original RS1 model, the bounds
in it are put on the set of two parameters: the ratio
$\kappa/\bar{M}_5$ and $m_1$ which is the mass of the lightest KK graviton.

We consider the diphoton production in the photon-induced process at
the LHC as a mean of looking for effects of low gravity scale
$\bar{M}_5$ in the Randall-Sundrum--like scenario with the small
curvature.

%%%%%%%%%%%%%%%%%%%%%%%%%%%%%%%%%%%%%%%%%%%%%%%%%%%%%%%%%%%%%%%%%%%%%

%%%%%%%%%%%%%%%%%%%%%%%
\section{Conclusions} %
%%%%%%%%%%%%%%%%%%%%%%%

With the forward detectors prepared by the ATLAS Forward Physics
Collaboration (AFP) and CMS-TOTEM Precision Proton Spectrometer
Collaboration (CT-PPS) \cite{afp1,afp2,afp3,totem}, it becomes
possible to investigate the exclusive photon-induced process $pp \to
p\gamma\gamma p \to p'Xp'$ (see Fig.~\ref{fig:sch}). In the present
paper we have studied the diphoton production $pp \to p \gamma
\gamma p \to p' \gamma \gamma p'$ at the LHC energy 14 TeV in the
framework of the Randall-Sundrum--like model with one warped ED and
small curvature of the 5-dimensional space-time. The consideration
was performed for two acceptance regions of the forward detector,
$0.015 < \xi < 0.15$ and $0.015 < \xi < 0.5$, where $\xi$ is the
fractional proton momentum loss of the incident protons.

The distributions in the photon transverse momenta $p_t$ with the
cut $|\eta| < 2.5$ imposed on the photon rapidity $\eta$ have been
calculated as a function of the reduced 5-dimensional Planck scale
$\bar{M}_5$ (see Figs.~\ref{fig:ptd15015_f} and
\ref{fig:ptd05015_f}). It was shown that the deviation from the SM
predictions gets higher as $p_t$ grows. The total cross sections
have been calculated for two acceptance regions depending on the cut
imposed on the transverse momenta of the final photon, $p_t
> p_{t,\min}$ (see Figs.~\ref{fig:ptcut15015_f} and
\ref{fig:ptcut05015_f}). Let us underline that in the RSSC model the
values of the cross sections don't depend on the curvature parameter
$\kappa$, provided $\kappa \ll \bar{M}_5$, what was satisfied in our
analysis. This allowed us to put the 95\% C.L. search limits for
$\bar{M}_5$ as a function of the integrated LHC luminosity (see
Figs.~\ref{fig:CL_pt300f} and \ref{fig:CL_pt500f}). For instance,
for $0.015 < \xi < 0.5$ and $p_t
> 300$ GeV, this limit for $\bar{M}_5$ is equal to 2.01 TeV, for the
integrated luminosity $L= 300$ fb$^{-1}$. For the HL-LHC integrated
luminosity $L= 3000$ fb$^{-1}$ and $p_t > 500$ GeV, we have found
that the 95\% C.L. search limit is equal to 2.93 TeV, for the same
acceptance region.  Any BSM scenario must be investigated in a
variety of processes in order to find the most appropriate one.
Recently the dimuon production in the photon-induced process at the
LHC was studied in \cite{Inan-Kisselev}, in which search limits for
$\bar{M}_5$ have been also obtained. The bounds on $\bar{M}_5$ in
the present article are better than the bounds in
\cite{Inan-Kisselev}.
%%%%%%%%%%%%%%%%%%%%%%%%%%%%%%%%%%%%%%%%%%%%%%%%%%%%%%%%%%%%%%%%%%%%%

%%%%%%%%%%%%%%
% References %
%%%%%%%%%%%%%%

%%%%%%%%%%%%%%%%%%%%%%%%%%%%%%%%%%%%%%%%%%%%%%%%%%%%%%%%%%%%%%%%%%%%%

%%%%%%%%%%%%%%%
\end{document}